\documentclass[format=sigconf,natbib=false,anonymous=false,screen]{acmart}

\def\myauthors{David Otero and Javier Parapar and Álvaro Barreiro}

\def\mytitle{Limitations of Automatic Relevance Assessments with Large Language Models for Fair and Reliable Retrieval Evaluation}

\AtBeginDocument{%
  }

\setcopyright{CC}
\acmBooktitle{Proceedings of the 48th International ACM SIGIR Conference on Research and Development in Information Retrieval (SIGIR '25), July 13--18, 2025, Padua, Italy}
\acmConference[SIGIR '25]{The 48th International ACM SIGIR Conference on Research and Development in Information Retrieval}{July 13--18, 2025}{Padua, Italy}


\usepackage{multirow}
\usepackage{soul}
\usepackage{subcaption}
\usepackage{amsmath}
\usepackage{graphicx}
\usepackage{url}
\usepackage[inline]{enumitem}
\usepackage{acronym}
\usepackage{booktabs}
\usepackage{longtable}
\usepackage{pdflscape}
\usepackage{tabularx}
\usepackage{afterpage}
\usepackage{siunitx}
\usepackage{algpseudocode, algorithm}
\usepackage{calc}
\usepackage[backend=biber,style=acmnumeric,datamodel=acmdatamodel,backref=true,backreffloats=false,sortcites]{biblatex}
\usepackage{todonotes}
\usepackage{datenumber}
\usepackage{colortbl}
\usepackage{cleveref}
\usepackage{pifont}

\hypersetup{
  pdfauthor={\myauthors},
  pdftitle={\mytitle},
  pdfsubject={Information Retrieval},
}

\crefname{enumi}{}{}

\newcommand{\ra}[1]{\renewcommand{\arraystretch}{#1}}

\DefineBibliographyStrings{american}{%
  backrefpage={cited on p.},
  backrefpages={cited on pp.}
}

\creflabelformat{equation}{#2\textup{#1}#3}

\author{David Otero}
\email{david.otero.freijeiro@udc.es}
\orcid{0000-0003-1139-0449}
\affiliation{
  \department[0]{IRlab, CITIC}
  \institution{Universidade da Coru\~{n}a}
  \city{A Coru\~{n}a}
  \country{Spain}
}

\author{Javier Parapar}
\email{javier.parapar@udc.es}
\orcid{0000-0002-5997-8252}
\affiliation{
  \department[0]{IRlab, CITIC}
  \institution{Universidade da Coru\~{n}a}
  \city{A Coru\~{n}a}
  \country{Spain}
}

\author{Álvaro Barreiro}
\email{barreiro@udc.es}
\orcid{0000-0002-6698-2946}
\affiliation{
  \department[0]{IRlab, CITIC}
  \institution{Universidade da Coru\~{n}a}
  \city{A Coru\~{n}a}
  \country{Spain}
}

\begin{document}

%%%%%%%%%%%%%%%%%%%%%%%%%%%%%%%%%%%%%%%%%%%%%%%%%
% Title
%%%%%%%%%%%%%%%%%%%%%%%%%%%%%%%%%%%%%%%%%%%%%%%%%
\title[Limitations of Automatic Relevance Assessments with LLMs for Fair and Reliable Retrieval Evaluation]{\mytitle}

%%%%%%%%%%%%%%%%%%%%%%%%%%%%%%%%%%%%%%%%%%%%%%%%%
% Abstract
%%%%%%%%%%%%%%%%%%%%%%%%%%%%%%%%%%%%%%%%%%%%%%%%%
\begin{abstract}

Offline evaluation of search systems depends on test collections. These benchmarks provide the researchers with a corpus of documents, topics and relevance judgements indicating which documents are relevant for each topic. While test collections are an integral part of Information Retrieval (IR) research, their creation involves significant efforts in manual annotation. Large language models (LLMs) are gaining much attention as tools for automatic relevance assessment. Recent research has shown that LLM-based assessments yield high systems ranking correlation with human-made judgements. These correlations are helpful in large-scale experiments but less informative if we want to focus on top-performing systems. Moreover, these correlations ignore whether and how LLM-based judgements impact the statistically significant differences among systems with respect to human assessments. In this work, we look at how LLM-generated judgements preserve ranking differences among top-performing systems and also how they preserve pairwise significance evaluation as human judgements. Our results show that LLM-based judgements are unfair at ranking top-performing systems. Moreover, we observe an exceedingly high rate of false positives regarding statistical differences.

\end{abstract}


\maketitle

%%%%%%%%%%%%%%%%%%%%%%%%%%%%%%%%%%%%%%%%%%%%%%%%%
% Sections
%%%%%%%%%%%%%%%%%%%%%%%%%%%%%%%%%%%%%%%%%%%%%%%%%
% !TeX spellcheck = en_GB

\section{Introduction}

Test collections allow for easy evaluation and comparison of Information Retrieval (IR) systems. A test collection consists of a set of relevance judgements (also called labels, assessments or qrels throughout this paper), which are annotations establishing the documents systems should retrieve for each user's need (also known as \textit{topics}). Collections built in venues like the Text REtrieval Conference (TREC) have enabled tremendous advancements in IR for many years. However, making the relevance judgements for a TREC-like test collection can be very expensive and labour-intensive. Low-budget collections could lead to biased and unstable evaluations.

Recently, several works explored the possibility of using large language models (LLMs) to assess the relevance of documents~\cite{Upadhyay2024a,Upadhyay2024b,Upadhyay2024c,Abbasiantaeb2024,Rahmani2024a,Thomas2024,MacAvaney2023,Faggioli2023}, since they are considerable cheaper than humans and show strong text-understanding capabilities. All these works use ranking correlations, such as Kendall's $\tau$~\cite{Kendall1948}, to validate the reliability of judgements generated with an LLM for IR evaluation. The rationale is that if judgements generated by an LLM can provide a similar ranking of systems to the ranking inferred from human judgements, the LLM could replace the human in terms of capturing run-level effectiveness~\cite{Upadhyay2024b}. Thus, the size of future collections in terms of judgements could increase drastically since the budget would no longer be an issue. The problem is that Kendall's $\tau$ is not very informative for evaluating agreement on the top-performing systems. Usually, the goal of developing a new system is surpassing the top of the leaderboard. Therefore, it is vital to distinguish meaningful differences between the best systems. Furthermore, some recent works have demonstrated that rank correlation is not enough~\cite{Otero2023b,Kutlu2018b}. Although judgements generated by an LLM may give good results in terms of system rankings, we know nothing about whether the results of a statistical test between any two systems would yield similar results. In addition, recent studies have highlighted that LLM-based judgements may suffer from other issues~\cite{Alaofi2024}.

In this paper, we aim to contribute to fill this gap. First, we have a closer look at how LLM-generated qrels are able to preserve the same ranking of systems as inferred from human judgements regarding top-performing systems. In particular, we use top-heavy rank correlations that penalize more misses at the top. Second, we evaluate how fair LLM-generated qrels are on a per-run basis. Last, we evaluate if significance decisions regarding the differences between evaluated systems under LLM judgements are similar to those observed under human judgements.

\textbf{The main highlights of our results are the following.} Correlations are not as good as they seem when we focus on the top-performing systems and degrade as we focus more and more on the first few systems. Additionally, there are many runs along the ranking that drop many positions when evaluated with LLM-based qrels, which suggests that they are not entirely fair. Finally, we observe that statistical tests made with LLM-based judgements yield an exceedingly high rate of false positives. Overall, our experiments suggest that LLMs are not yet ready to replace humans for making gold relevance assessments in IR offline evaluation.

% !TeX spellcheck = en_GB

\section{Differences among Top Systems}

In this section, we have a closer look at how LLM-based qrels are able to preserve the same ranking of systems among top-performing ones. We used the data of the SynDL dataset~\cite{Rahmani2024b} to perform our experiments. This dataset is based on the passage ranking task of the Deep Learning Track, from years 2019 to 2023. Each year, track organizers created a set of queries from which some received human judgements. For example, in DL19, only 43 out of the 200 queries received expert assessments. The creators of SynDL used GPT-4 to provide labels for the rest of the queries that did not receive human judgements. For more details about how the LLM-based labels have been generated, we refer the reader to the paper~\cite{Rahmani2024b}. \Cref{tab:data} summarises the data we have used. It is important to note that the set of topics that have human judgements and the set of topics that have LLM judgements are disjoint.

\begin{table}
  \centering
  \footnotesize
  \caption{Statistics of the datasets used for experimentation. Note that $T_{H} \cap T_{LLM} = \emptyset$.}
  \label{tab:data}
  \setlength\tabcolsep{2pt}
  \ra{1.0}
  \begin{tabular}{@{}lcrrrrr@{}}
    \toprule
    \textbf{Data} && \textbf{DL-19} & \textbf{DL-20} & \textbf{DL-21} & \textbf{DL-22} & \textbf{DL-23}\\
    \midrule
    Topics with human judgements ($\lvert T_{H} \rvert$) && 43 & 54 & 53 & 76 & 82\\
    Topics with LLM judgements ($\lvert T_{LLM} \rvert$) && 157 & 146 & 424 & 424 & 618\\
    \arrayrulecolor{gray!30}
    \midrule
    \arrayrulecolor{black}
    Average judgements per topic in $T_{H}$ && 215 & 211 & 204 & 5084 & 272\\
    Average judgements per topic in $T_{LLM}$ && 70 & 128 & 157 & 133 & 73\\
    \arrayrulecolor{gray!30}
    \midrule
    \arrayrulecolor{black}
    Submissions && 36 & 59 & 63 & 100 & 35 \\
    Pairs of submissions && 630 & 1711 & 1953 & 4950 &595 \\
    \bottomrule
  \end{tabular}
\end{table}

To evaluate to which extent LLM-based qrels can preserve the ordering among top-performing systems, we have computed two different top-weighted correlations, namely $\tau_{AP}$~\cite{Yilmaz2008} and Rank-Biased Overlap (RBO)\footnote{We used the parameter $p$ = 0.7 to focus on the first third or four runs~\cite[Section 4.5]{Thomas2024}}~\cite{Webber2010}. The main idea is that, when comparing two rankings of systems, discrepancies among those systems with high positions are penalized more than those with low positions. \Cref{tab:correlation} shows the top-weighted systems ranking correlations, along with Kendall's $\tau$, between official TREC judgements and LLM-based judgements. As already demonstrated in previous work~\cite{Rahmani2024b,Upadhyay2024b,Faggioli2023}, LLM-based qrels show competitive values regarding Kendall's $\tau$. However, we can observe that the correlations that penalize more for failures in higher positions give significantly lower values, which highlights the idea that the judgements generated with an LLM are not able to identify the best systems.

%%%%%%%%%%%%%%%%%%%%%%%%%%%%%%%%%%%%%%%%%%
% Correlations
%%%%%%%%%%%%%%%%%%%%%%%%%%%%%%%%%%%%%%%%%%
\begin{table}
  \centering
  \small
  \caption{Systems Ranking correlations between official TREC human-made judgements and LLM-based judgements.  }
  \label{tab:correlation}
  \setlength\tabcolsep{2pt}
  \ra{1.0}
  \begin{tabular}{@{}lllrrrrr@{}}
    \toprule
    \multirow{2.3}{*}{\textbf{Metric}} &&& \multicolumn{5}{c}{\textbf{Dataset}} \\
    \cmidrule{4-8}
    &&&\multicolumn{1}{c}{DL-2019}&\multicolumn{1}{c}{DL-2020}&\multicolumn{1}{c}{DL-2021}&\multicolumn{1}{c}{DL-2022}&\multicolumn{1}{c}{DL-2023}\\
    \midrule
    \multirow{3}{*}{\textbf{AP}}& $\tau$ && 0.62 & 0.85 & 0.80 & 0.75 & 0.79 \\
    & $\tau_{AP}$ && 0.49 & 0.70 & 0.72 & 0.63 & 0.66 \\
    & $RBO~(p = 0.7)$ && 0.11 & 0.43 & 0.52 & 0.51 & 0.33  \\
    \midrule
    \multirow{3}{*}{\textbf{NDCG}}& $\tau$ && 0.80 & 0.90 & 0.87 & 0.85 & 0.88 \\
    & $\tau_{AP}$ && 0.79 & 0.84 & 0.85 & 0.77 & 0.83 \\
    & RBO ($p$ = 0.7) && 0.93 & 0.54 & 0.97 & 0.58 & 0.69  \\
    \bottomrule
  \end{tabular}
\end{table}

\subsection{Fairness}

These correlations are useful for showing overall performance on large-scale experiments, but they hide the fact that some runs may be treated differently in the two evaluations. That is, it could be the case that, although correlations are actually high, some runs are losing (or gaining) many positions in the ranking.

\begin{figure}
  \centering
  \includegraphics[width=\linewidth]{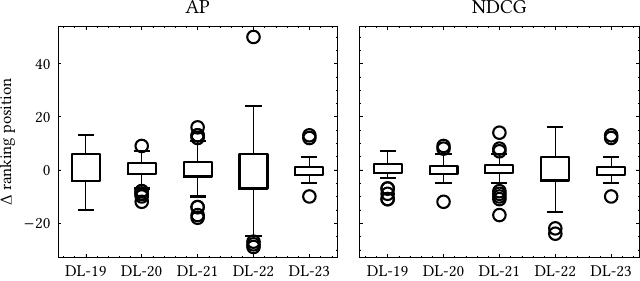}
  \caption{Results of the fairness experiment. For each year and metric, we plot the distribution of the changes in ranking positions that runs undergo when evaluated with LLM-generated assessments. A positive change means that the run is ranked at a higher position under LLM judgements, while a negative value means that the run is ranked at a lower position.}
  \label{fig:fairness}
\end{figure}

Following a common approach in past work~\cite{Voorhees2018,Otero2023a}, we evaluate the \textbf{fairness}. When evaluated with LLM-generated judgements, we compute the change that each run suffers with respect to the ranking of systems inferred from human judgements. We show the results of this experiment in \Cref{fig:fairness}. This figure depicts the distributions, as boxplots, of the changes in ranking positions that runs undergo when evaluated with LLM-based assessments. A positive value means that the run is at a higher position with LLM qrels, while a negative value means that the run is at a lower position. Overall, we observe that almost all runs suffer some change in position when evaluated with LLM judgments. More worrying is the fact that there are, in all cases, regardless of the metric or the collection, runs that suffer changes of 5 or more positions, even reaching changes of more than 40 positions.

%%%%%%%%%%%%%%%%%%%%%%%%%%%%%%%%%%%%%%%%%%
% Ranking Drops
%%%%%%%%%%%%%%%%%%%%%%%%%%%%%%%%%%%%%%%%%%

\begin{figure}
  \centering
  \includegraphics[width=\linewidth]{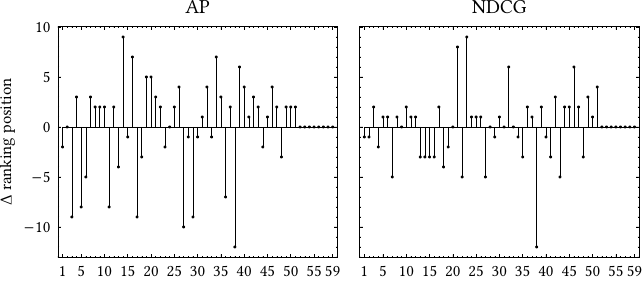}
  \caption{Per-run drops in DL-2020. The x-axis shows the position of each run in the ranking inferred from human judgements. The y-axis shows the change in ranking position when evaluated with LLM-based judgements. A positive change means that the run is ranked at a higher position under LLM judgements, while a negative value means that the run is ranked at a lower position.}
  \label{fig:drop-runs}
\end{figure}

As a follow-up analysis, in \Cref{fig:drop-runs} we include the change suffered by each of the runs that participated in the DL-2020. The x-axis shows the ranking of the runs inferred from human judgements, while the y-axis shows the change that each run suffers, in terms of ranking position, when evaluated with LLM judgements. Overall, we observe that most of the runs suffer a change. Only 10 and 13 runs stay at the same position when evaluated with AP and NDCG, respectively. In other words, 83\% of runs with AP, and 78\% with NDCG, get a different position when evaluated with LLM-based qrels.

% !TeX spellcheck = en_GB

\section{Preservation of Statistical Significances}

In the previous section, we evaluated biases exhibited by LLM-based relevance assessments in ranking correlations and ranking swaps, especially among top-performing systems. In this section, we study the reliability of LLM-generated judgements to evaluate retrieval systems from the perspective of statistical tests. Assuming that human judgements are the ground truth, we evaluate whether the LLM labels can provide a similar pairwise significance evaluation, i.e., whether decisions regarding the statistical difference between a pair of systems are the same using human labels and LLM labels.

Let $Q_{H}$ be the judgements made by human assessors over a set $T_{H}$ of topics, and let $Q_{LLM}$ be the judgements made by an LLM over a set $T_{LLM}$ of topics. We do not assume any overlap between $T_{H}$ and $T_{LLM}$, in fact, they may be completely disjoint sets and have different sizes. Let $R$ be the set of runs under evaluation, where each run includes results for all topics $T_{H} \cup T_{LLM}$. Using an evaluation measure of choice (e.g. Average Precision), we compute per-topic scores for each run in $R$ using both sets of judgements independently. Thus, we have two arrays of per-topic scores for each run, one obtained using $Q_{H}$ and the other obtained with $Q_{LLM}$. With these per-topic scores, we compute a $p$-value for each pair of runs,\footnote{We use a two-sided test.} for both sets of qrels. So, in the end, we have, on the one hand, a $p$-value for each pair of runs computed from the scores obtained with $Q_{H}$ and, on the other hand, a different $p$-value for each pair of runs computed from the scores obtained with $Q_{LLM}$. Thus, for each pair of runs, given a significance level $\alpha$, we know if they are significantly different (or not) under $Q_{H}$ and under  $Q_{LLM}$. Now, assuming that the significance decisions derived from the $Q_{H}$ qrels are the ground truth, we evaluate how the decisions derived from $Q_{LLM}$ match the former. In particular, we compute the true/false positives/negatives:

\begin{itemize}[leftmargin=*,itemsep=5pt]
  \item \textbf{True Positives (TP)}: these are the cases where both set of judgements marked a significant result.

  \item \textbf{False Negatives (FN)}: these are the cases where a pair of runs is marked as significantly different under $Q_{H}$, but not under $Q_{LLM}$.

  \item \textbf{True Negatives (TN)}: these are the cases where both set of judgements marked a non-significant result.

  \item \textbf{False Positives (FP)}: these are the cases where a pair of runs is marked as non-significant under $Q_{H}$, but marked as significant under $Q_{LLM}$.
\end{itemize}

\noindent We report these metrics as percentages over the gold counts of significant and non-significant. TP will be the percentage of correctly detected positives over the total of gold positives, FN will be the ratio of incorrectly detected negatives over the total of positives, and so on. Thus, $TP + FN = 100\%$ and $TN + FP = 100\%$.

\subsection{Experimental Setup}

We used both Average Precision (AP) and NDCG (Normalized Discounted Cumulative Gain) with a 1000 cutoff for scoring the runs. To compute the $p$-values, we employed a two-sided Wilcoxon Signed-Rank test as recommended in previous literature~\cite{Parapar2021,Otero2025}, and set $\alpha = 0.05$.

\subsection{Results}

%%%%%%%%%%%%%%%%%%%%%%%%%%%%%%%%%%%%%%%%%%
% True/False Positives/Negatives
%%%%%%%%%%%%%%%%%%%%%%%%%%%%%%%%%%%%%%%%%%
\begin{table}
  \centering
  \small
  \caption{Classification results observed using the synthetic qrels on different datasets with AP and NDCG.}
  \label{tab:agreements}
  \setlength\tabcolsep{2pt}
  \ra{1.0}
  \begin{tabular}{@{}lllrrrrr@{}}
    \toprule
    \multirow{2.3}{*}{\textbf{Metric}} &&& \multicolumn{5}{c}{\textbf{Dataset}} \\
    \cmidrule{4-8}
    &&&\multicolumn{1}{c}{DL-2019}&\multicolumn{1}{c}{DL-2020}&\multicolumn{1}{c}{DL-2021}&\multicolumn{1}{c}{DL-2022}&\multicolumn{1}{c}{DL-2023}\\
    \midrule
    \multirow{4.3}{*}{\textbf{AP}}& TP && 80\% & 94\% & 96\% & 95\% & 97\% \\
                                  & FN && 20\% &  6\% &  4\% &  5\% &  3\% \\
    \arrayrulecolor{gray!30}
    \cmidrule{2-8}
    \arrayrulecolor{black}
                                  & TN && 29\% & 42\% & 17\% & 20\% & 11\%  \\
                                  & FP && 71\% & 58\% & 83\% & 80\% & 89\% \\
    \midrule

    \multirow{4.3}{*}{\textbf{NDCG}}& TP && 92\% & 98\% & 98\% & 97\% & 98\% \\
                                    & FN &&   8\% &  2\% &  2\% &  3\% &  2\% \\
    \arrayrulecolor{gray!30}
    \cmidrule{2-8}
    \arrayrulecolor{black}
                                    & TN && 33\% & 46\% & 28\% & 26\% &  6\% \\
                                    & FP && 67\% & 54\% & 72\% & 74\% & 94\%  \\
    \bottomrule
  \end{tabular}
\end{table}

We show the results of our experiments in \Cref{tab:agreements}. This table reports the TP/TN/FP/FN percentages both when scoring the runs with AP (upper half) and NDCG (lower half).

From this table, we highlight several interesting results. First, we see systematically better figures with NDCG than MAP---higher positives, lower negatives---, suggesting that NDCG is more robust for evaluating systems with LLM-qrels. We observe that LLM-generated labels detect most of the existing significant differences, since TP ratios are near 100\% almost in every case. This result holds independently of using AP or NDCG, although TP figures obtained with NDCG are systematically higher. Interestingly, we observe a high number of false positives for every year and metric. These cases represent situations were the LLM judgements $Q_{LLM}$ marked a significant difference when there was none. In other words, some pairs of systems are not significantly different under $Q_{H}$ but they are under $Q_{LLM}$. It is a concerning result, since judgements created with an LLM would lead a researcher to incorrectly conclude and present a supposedly significant result. However, we hypothesize that this might be happening because of the difference in sample sizes between both evaluations. As observed in \Cref{tab:data}, the number of topics that include LLM-generated labels is much higher than the topics that have human judgements. It is known that more topics provide more power when performing a statistical test~\cite{Carterette2008,Sanderson2005,Webber2008}. Therefore, these false positives could be happening for this reason.

To check this hypothesis, we sampled random subsets of $T_{LLM}$ so that $\lvert T_{H} \rvert = \lvert T_{LLM} \rvert$. In this way, both evaluations have an equal sample size. We sampled 200 times, and computed the average of ratios over these iterations. We present the results in \Cref{tab:agreements-sampling}.

%%%%%%%%%%%%%%%%%%%%%%%%%%%%%%%%%%%%%%%%%%
% True/False Positives/Negatives SAMPLING
%%%%%%%%%%%%%%%%%%%%%%%%%%%%%%%%%%%%%%%%%%
\begin{table}
  \centering
  \small
  \caption{Classification results observed using the SynDL synthetic qrels on different datasets with AP and NDCG. We randomly undersample $T_{LLM}$ topics such that $\lvert T_{LLM} \rvert = \lvert T_{H} \rvert$.}
  \label{tab:agreements-sampling}
  \setlength\tabcolsep{2pt}
  \ra{1.0}
  \begin{tabular}{@{}lllrrrrr@{}}
    \toprule
    \multirow{2.3}{*}{\textbf{Metric}} &&& \multicolumn{5}{c}{\textbf{Dataset}} \\
    \cmidrule{4-8}
    &&&\multicolumn{1}{c}{DL-2019}&\multicolumn{1}{c}{DL-2020}&\multicolumn{1}{c}{DL-2021}&\multicolumn{1}{c}{DL-2022}&\multicolumn{1}{c}{DL-2023}\\
    \midrule
    \multirow{4.3}{*}{\textbf{AP}}   & TP && 68\% & 88\% & 88\% & 89\% & 91\% \\
                                     & FN && 32\% & 12\% & 12\% & 11\% &  9\%  \\
        \arrayrulecolor{gray!30}
    \cmidrule{2-8}
    \arrayrulecolor{black}
                                     & TN && 57\% & 62\% & 50\% & 51\% & 28\% \\
                                     & FP && 43\% & 38\% & 50\% & 49\% & 72\% \\
    \midrule

    \multirow{4.3}{*}{\textbf{NDCG}} & TP && 83\% & 94\% & 91\% & 92\% & 93\% \\
                                     & FN && 17\% &  6\% &  9\% &  8\% &  7\% \\
    \arrayrulecolor{gray!30}
    \cmidrule{2-8}
    \arrayrulecolor{black}
                                     & TN && 58\% & 63\% & 64\% & 55\% & 33\%\\
                                     & FP && 42\% & 37\% & 36\% & 45\% & 67\% \\
    \bottomrule
  \end{tabular}
\end{table}

These results show that, indeed, under equal sample size conditions, the number of false positives is lower, supporting our previous hypothesis, although they are still concerning. We also observe lower values of true positives. Now some differences that are real under $Q_{H}$ are being ignored under $Q_{LLM}$. Leaving some real advancements behind can also hinder progress in the field. Regarding the differences between true negatives and true positives, we see that LLM-generated judgements are equally effective at detecting true significances and non-significances. In this case, we also observe better figures with NDCG, a result in line with previous research in related fields~\cite{Valcarce2018c}. Overall, the results of this experiment showcase that LLMs are not yet capable of producing IR relevance labels that preserve the same statistical differences as human labels.

Similar to the previous section, we are also interested in whether a run has more or less significant differences when evaluated under the LLM-based qrels. For every run, we count the significant differences observed under human and LLM judgements against this run. Then, a \emph{drop} happens when the number is higher in the first case, meaning that a run is losing some significances under the LLM labels. We also undersampled $T_{LLM}$ 50 times and averaged the results. We plot the distribution of these drops in \Cref{fig:fairness}.

\begin{figure}
  \centering
  \includegraphics[width=\linewidth]{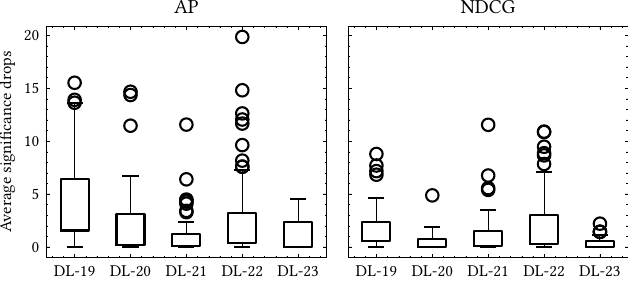}
  \caption{Distribution of the drops in significance counts of the runs when evaluated with LLM labels. Each value is the average over 50 iterations of undersampling $T_{LLM}$.}
  \label{fig:drops}
\end{figure}

This figure shows the distribution of the per-run significance drops, averaged over 50 undersampling iterations, for the different year and effectiveness measures we have used. The ideal case here would be observing narrow boxes near 0 without outliers. This would mean that, on average, no run is treated differently in terms of significances. However, we observe that, to a greater or lesser degree, there is always some run treated differently, regardless of the year and the effectiveness measure. This suggests that the false negatives we observed in a previous experiment (see \Cref{tab:agreements-sampling}) are concentrated within just a few runs. In other words, the false negatives appear because the LLM qrels consistently fail to correctly evaluate the same few runs.

% !TeX spellcheck = en_GB

\section{Conclusions}

Making relevance judgements is the most resource-intensive task when creating new collections for IR evaluation. Recent research has explored the possibility of using LLMs to assess the relevance of documents, showing promising results: LLMs can produce accurate labels compared to human assessments and reliable rankings of systems. However, little work has evaluated the validity of these judgements in terms of distinguishing the best top-performing systems and preserving statistical significance stability. In this work, we looked at these two issues and evaluated the reliability of LLM labels for IR evaluation from different perspectives. First, we evaluated if LLM-based qrels were able to correctly rank top-performing systems by employing top-heavy ranking correlation. Results showed that correlations degrade as we focus more and more on the best systems. We also evaluated whether significance decisions regarding the differences between evaluated systems are similar to those observed under human judgements. Our results showed that LLM-generated judgements yield an exceedingly high rate of false positives. This result suggests they are biased to incorrectly mark a significant difference when there is none. We also looked at whether any of the runs suffered great changes in their ranking position when evaluated with LLM judgments. In this experiment, we observed that there are runs that suffer drops of up to 50 positions, which suggests that these LLM-generated judgements are not entirely fair for evaluating IR systems.

\begin{acks}

\sloppy The authors thank the financial support supplied by the grant PID2022-137061OB-C21 funded by MICIU/AEI/10.13039/501100011033 and by “ERDF/EU”. The authors also thank the funding supplied by the Consellería de Cultura, Educación, Formación Profesional e Universidades (accreditations ED431G 2023/01 and ED431C 2025/49) and the European Regional Development Fund, which acknowledges the CITIC Research Center as a Center of Excellence and recognizes it as a Member of the CIGUS Network for the period 2024-2027.

\end{acks}

%%%%%%%%%%%%%%%%%%%%%%%%%%%%%%%%%%%%%%%%%%%%%%%%%
% References
%%%%%%%%%%%%%%%%%%%%%%%%%%%%%%%%%%%%%%%%%%%%%%%%%
\printbibliography

\end{document}